\documentclass[sigconf, authorversion=true, nonacm=true]{acmart}

\AtBeginDocument{%
  \providecommand\BibTeX{{%
    \normalfont B\kern-0.5em{\scshape i\kern-0.25em b}\kern-0.8em\TeX}}}

\begin{document}

\title{Tutorial on Deep Learning for Human Activity Recognition}

\author{Marius Bock}
\affiliation{%
  \institution{University of Siegen}
  \streetaddress{H\"olderlinstr. 3}
  \postcode{57076}
  \country{Germany}
}
\email{marius.bock@uni-siegen.de}

\author{Alexander H\"olzemann}
\affiliation{%
  \institution{University of Siegen}
  \streetaddress{H\"olderlinstr. 3}
  \postcode{57076}
  \country{Germany}
}
\email{alexander.hoelzemann@uni-siegen.de}

\author{Michael Moeller}
\affiliation{%
  \institution{University of Siegen}
  \streetaddress{H\"olderlinstr. 3}
  \postcode{57076}
  \country{Germany}
}
\email{michael.moeller@uni-siegen.de}

\author{Kristof Van Laerhoven}
\orcid{0000-0001-5296-5347}
\affiliation{%
  \institution{University of Siegen}
  \streetaddress{H\"olderlinstr. 3}
  \postcode{57076}
  \country{Germany}
}
\email{kvl@eti.uni-siegen.de}
\renewcommand{\shortauthors}{Bock, H\"olzemann, et al.}

\begin{abstract}
  Activity recognition systems that are capable of estimating human activities from wearable inertial sensors have come a long way in the past decades. Not only have state-of-the-art methods moved away from feature engineering and have fully adopted end-to-end deep learning approaches, best practices for setting up experiments, preparing datasets, and validating activity recognition approaches have similarly evolved.
  This tutorial was first held at the 2021 ACM International Symposium on Wearable Computers (ISWC’21) and International Joint Conference on Pervasive and Ubiquitous Computing (UbiComp’21). The tutorial, after a short introduction in the research field of activity recognition, provides a hands-on and interactive walk-through of the most important steps in the data pipeline for the deep learning of human activities.
  
  All presentation slides shown during the tutorial, which also contain links to all code exercises, as well as the link of the GitHub page of the tutorial can be found on:\newline \textbf{\url{https://mariusbock.github.io/dl-for-har}}
\end{abstract}


\maketitle

\section*{Intended Audience}
We do not require prior knowledge in human activity recognition or deep learning techniques, but instead will demonstrate basic concepts, best practices and upcoming techniques to create human activity recognition systems with deep learning approaches. We assume that participants of this tutorial are already familiar with the basics of Python as a programming language\footnote{For those who would like to prepare, we highly recommend this online tutorial on Python for beginners linked at: \url{https://www.python.org/about/gettingstarted/}}.
The focus of this tutorial is on the correct way to establish a data pipeline for human activity recognition from inertial data: Using existing public activity datasets, participants are shown how to systematically design and evaluate deep learning architectures for activity recognition.

This tutorial is designed to function as a guide in the typical processes to design and evaluate deep learning architectures, once an activity recognition dataset has been recorded and annotated. We show through code examples in a step-by-step process why certain steps are needed, how they affect the system's outcome, and which pitfalls present themselves when designing a deep learning classifier for activity recognition. 
\vspace{8mm}




\begin{figure*}[ht]
  \centering
  \includegraphics[width = 0.999\linewidth]{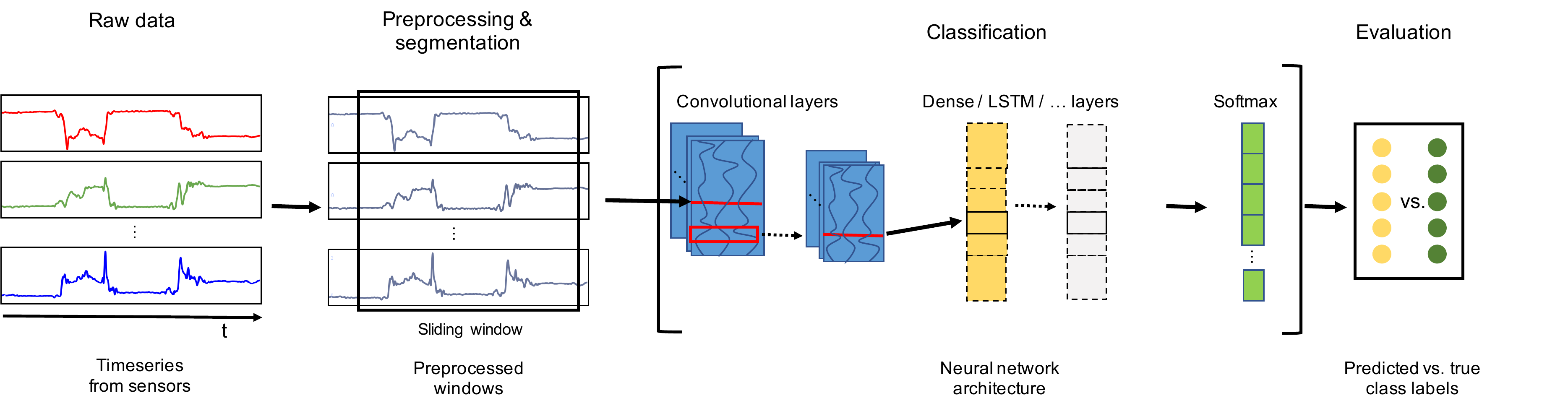}
  \caption{This updated version of the ARC as proposed by Bulling et al. \cite{bullingTutorialHumanActivity2014}, called Deep Learning Activity Recognition Chain (DL-ARC), shows the different phases of this tutorial. Given a set of inertial sensors, the raw data is first preprocessed and segmented into sliding windows. Unlike the ARC, feature extraction becomes obsolete in the DL-ARC. The raw, segmented and preprocessed sensor data is fed into a classification neural network to obtain a set of predicted labels. The obtained set of predicted labels is assessed using state-of-the-art evaluation metrics.}
  \Description{Illustration of the DL-ARC pipeline incorporating highlighting all pipeline steps as well as the non-necessity of the feature extraction step which was part of the original ARC pipeline as proposed by Bulling et al. \cite{bullingTutorialHumanActivity2014}}
  \label{fig:dl-arc}
\end{figure*}

\section{Motivation}
Physical activities constitute pivotal information that allows us to structure our daily lives. Knowing when which activities happen, and how they are performed, can reveal much about persons' intentions, habits, fitness, and state of mind; It is therefore not surprising that researchers into Ubiquitous Computing and Wearable Computers have displayed a growing interest in the machine recognition of human activities, also known as Human Activity Recognition (often abbreviated as HAR).

In 2014, Bulling et al. \cite{bullingTutorialHumanActivity2014} designed and organized an exceptionally well-received tutorial on human activity recognition from wearable sensor data. Within their tutorial, they introduced concepts such as the Activity Recognition Chain (ARC), a framework for designing and evaluating activity recognition systems, as well as a case study demonstrating how to work with this ARC. Since the release of said tutorial, a lot has happened in the area of human activity recognition research. Within the last decade, deep learning methods have shown to outperform classical Machine Learning algorithms (see for instance \cite{krizhevskyImagenetClassificationDeep2012, collobertNaturalLanguageProcessing2011, szegedyGoingDeeperConvolutions2015} for a few examples) and, as a product of this success, have led to studies investigating the effectiveness of deep learning in activity recognition (see for example  \cite{ordonezDeepConvolutionalLSTM2016, hammerlaDeepConvolutionalRecurrent2016}). 

The ARC, as proposed by Bulling et al. \cite{bullingTutorialHumanActivity2014}, did not yet include deep learning techniques and was solely based on classical Machine Learning approaches. With studies having demonstrated the effectiveness of deep learning for activity recognition with wearable sensors, we argue that releasing an updated tutorial that is adapted to work with deep learning techniques was long overdue. Within this tutorial we therefore introduced the Deep Learning Activity Recognition Chain (DL-ARC), which is a reworked version of the original ARC, encompassing the advances that have been made over the years within the field of deep learning for human activity recognition and deep learning in general. Our work directly ties into the works of Bulling et al. \cite{bullingTutorialHumanActivity2014} and functions as a step-by-step framework to apply deep learning to any activity recognition use case. Within this tutorial, we will show how state-of-the-art models can be achieved, while along the way explaining all design choices in detail.

Beyond introducing deep learning approaches for activity recognition, other advances within the deep learning community from the past years, such as modifications to the loss function and data augmentation strategies, etc. will be discussed and analysed for their impact via comparison of results with an without employing said modifications.

\section{Outline}

This tutorial is targeted towards anyone who wants to understand how to successfully establish a data pipeline for HAR from inertial data. Participants of this tutorial are shown how to systematically design and evaluate deep learning architectures for activity recognition using our suggested pipeline, namely the DL-ARC. Figure \ref{fig:dl-arc} illustrates the pipeline, which most prominently differs from the original ARC in that it does not contain a separate feature engineering step. As predictive performance of classical Machine Learning approaches highly relies on sophisticated, handcrafted features \cite{pouyanfarSurveyDeepLearning2018}, Bulling et al. \cite{bullingTutorialHumanActivity2014} included a feature extraction phase within the original ARC. However, one main advantage of using deep learning techniques is that they are able to automatically extract discriminative features from raw data input \cite{najafabadiDeepLearningApplications2015}, thus making the feature extraction pipeline step obsolete.

To successfully understand the contents of this tutorial, we do not require participants to have prior knowledge in HAR or deep learning techniques. Participants are be able to follow along the content of the tutorial through a GitHub page. The tutorial is structured into three main phases. The first part will function as an introduction to the tutorial, the research field of human activity recognition as well as the structure of the tutorial. Next, the second part, will introduce the DL-ARC through a mix of theoretical and hands-on coding exercises. Lastly, the third part, will demonstrate next possible research steps by introducing advances within the deep learning community, which though applied in different research fields (e.g. computer vision) hold the potential to improve the predictive performance of the pipeline. 

In the following, you find a compiled list of all questions which will be answered within each phase:
 
\vspace{0.3cm}
\noindent \textbf{Introduction \ \ }
\vspace{-0.1cm}
    \begin{itemize}
        \item Why this tutorial? What is its goal? What can participants expect?
        \item How is the GitHub page of this tutorial structured? 
        \item What is Human Activity Recognition? Where does it find its applications? How was it approached in the past?
        \item What are advantages of deep learning based approaches compared to classical approaches?
        \item What are sample deep learning architectures for HAR? What is the DeepConvLSTM?
    \end{itemize}
    
\vspace{0.1cm}
\noindent \textbf{The DL-ARC pipeline \ \ }
\vspace{-0.1cm}
    \paragraph{Data collection and data analysis}
        \begin{itemize}
                \item How is data collection usually performed in the context of HAR? 
                \item How does one typically analyse HAR datasets to get a better intuition of the use case at hand?
                \item What is the RealWorld HAR (RWHAR) dataset \cite{sztylerOnBodyLocalizationWearable2016}?
        \end{itemize}
    \paragraph{Preprocessing}
        \begin{itemize}
            \item What are common issues when dealing with sensor datasets?
            \item How does one deal with missing data in a sensor dataset?
            \item How does one deal with varying sampling rates and sensitives of sensors? What is resampling? What is scaling/ normalization? 
            \item What is the sliding window approach? Why do we need it?
        \end{itemize}
    \paragraph{Network architecture and training} 
        \begin{itemize}
            \item What is deep learning? Of what parts does a typical (deep learning) training loop consist of?
            \item How to assess the predictive performance of a trained classification model? What are common metrics?
            \item What is the DeepConvLSTM \cite{ordonezDeepConvolutionalLSTM2016}? What are the main parts of the architecture and what are their purpose?
            \item How does one train a sample deep learning model using our GitHub repository?
        \end{itemize}
    \paragraph{Validation and testing}
        \begin{itemize}
            \item Why do we need validation and testing? What is overfitting?
            \item Which different validation methods are typically used within the field of HAR?
            \item What is hypertuning? How should it be applied?
        \end{itemize}
\pagebreak
\noindent \textbf{Next research steps \ \ }
    \begin{itemize}
        \item What are examples of recent advances in the field of deep learning for computer vision? 
        \item Can their advantageous behavior carry over to HAR and lead to an improved predictive performance of the classification network?
    \end{itemize}

\section{Related work}
Recent advances in human activity recognition has empowered further development in fields like medical computer science \cite{horiInertialMeasurementUnitBased2020, horstUnderstandingInterpretationMachine2019}, industrial manufacturing \cite{grzeszickDeepNeuralNetwork2017, koskimakiActivityRecognitionUsing2009}, smart environments \cite{kockemannOpenSourceDataCollection2020, gilmanInternetThingsSmart2020} and sports analysis \cite{holzemannUsingWristWornActivity2018, maNPURGBDDataset2021}. However, some of the important basics as introduced in \cite{bullingTutorialHumanActivity2014}, on which much research in this area has built on, are not state of the art anymore. At the latest since deep learning has found its way into activity recognition, the standards of former algorithms are often no longer completely applicable. 

Feature engineering, as known from classical machine learning approaches, is no longer necessary since \cite{hammerlaPDDiseaseState2015} showed that raw sensor data can be processed by using deep neural networks, even though, recent publications like \cite{chenImprovingHumanActivity2021} show that it can be beneficial when it comes to specific domains and circumstances or when using specific architectures of neural networks \cite{dziezycCanWeDitch2020}. In classical machine learning, however, statistics of the first order are mostly used (mean, variance, median values, etc.). Unlike, when feature engineering is applied in the context of deep learning, the calculated features are usually of higher order. 

New technologies, such as transfer learning \cite{hoelzemannDiggingDeeperBetter2020, gjoreskiCrossDatasetDeepTransfer2019, qinCrossDatasetActivityRecognition2019, duTransferLearningHuman2019}, data augmentation \cite{hoelzemannDataAugmentationStrategies2021, liActivityGANGenerativeAdversarial2020, rashidTimesSeriesDataAugmentation2019} and active learning \cite{hoqueAALOActivityRecognition2012}, but also the constant search for more efficient network architectures \cite{ordonezDeepConvolutionalLSTM2016, murahariAttentionModelsHuman2018, gjoreskiClassicalDeepLearning2020} are becoming an important part in the human activity recognition community and are therefore constantly driving research in the field of deep learning. 
Two types of human activity recognition can be distinguished: (1) video data based, e.g. \cite{yanSpatialTemporalGraph2018} and \cite{souriFastWeaklySupervised2021} and (2) inertial sensor data based activity recognition. An inertial sensor consists of at least an accelerometer and a gyroscope, but is often supplemented by a magnetometer. These datasets recorded with such sensors can be further enriched with additional information by additional sensor technology, such as temperature or light sensors. Combining an accelerometer and gyroscope alone can already significantly improve the classification results of machine learning models \cite{guoWearableSensorBased2016}.

Therefore, current scientific research also focuses on fusion techniques for multimodal datasets, e.g. \cite{maleRecognitionHumanActivity2021} and \cite{munznerCNNBasedSensorFusion2017}, or developing network architectures that are capable to handle both input types of data \cite{trumbleTotalCapture3D2017, henschelAccurateLongTermMultiple2020}.
However, algorithms from computer graphics and computer vision or even language processing often serve as inspiration, which are then transferred to sensor data-based human activity recognition. This results in a gap between both deep learning disciplines, in which the state of the art of the sensor-based human activity recognition mostly lacks behind the computer vision.

Nevertheless, the number of publications related to these topics increased constantly throughout the years (see \cite{hernandezLiteratureReviewTransfer2020} and \cite{rajspSystematicLiteratureReview2020} for an overview), but the focus on important fundamentals seems to be lost. These issues include that some works are difficult to reproduce as they lack open source code, focus on very specific aspects and/ or do not generalize well beyond the data set they have been developed on.


\section{Format and detailed schedule}
This tutorial was originally designed to be a full-day tutorial, taking about 6 hours plus breaks to work through. Given this limited timeframe we presented participants a less extensive version of the tutorial and left certain parts as take-home assignments. Nevertheless, our GitHub page contains all necessary information and hints so that participants can repeat all contents of the tutorial at any point in the future. The GitHub page guides participants through a mix of theoretical and hands-on exercise sessions. The two types of sessions will be either provided in the form of information directly displayed on the website via a slide show presentation or via Jupyter notebooks accessed through Google Colab\footnote{See \url{https://colab.research.google.com}}. 

Within their tutorial, Bulling et al. \cite{bullingTutorialHumanActivity2014} introduced the ARC. With our tutorial we reworked the ARC to be suited for deep learning and propose a new version of it called the DL-ARC. As previously mentioned we structured the tutorial into three main phases. 

\vspace{0.1cm}
\noindent \textbf{Introduction \ \ } The first phase gives an overview of the tutorial and an introduction to human activity recognition (HAR) and deep learning. We start by explaining the purpose of the tutorial, how it compares to the work of Bulling et al. \cite{bullingTutorialHumanActivity2014}, and what they can expect to learn within the next hours, as well as introduce our GitHub page which contains all the content covered within the tutorial. Afterwards, we will have a longer session where we will cover necessary theoretical knowledge. We will first answer basic questions like what HAR is and how it was approached in the past. To do so, we will sketch a timeline which will give participants a brief understanding how the area of HAR developed over the years. Next, we will go over explaining basic concepts of deep learning and where its advantages lie compared to classical (Machine Learning) approaches. We will then extend the timeline previously sketched with the influence of deep learning on the field of HAR, highlighting the most important studies conducted in the field, which demonstrated the effectiveness and applicability of deep learning for HAR, as well as other notable milestones in the other fields. 

With the brief introduction of related work within the field of deep learning for HAR, we will also introduce the sample neural network architecture which participants will use throughout the tutorial, namely the DeepConvLSTM \cite{ordonezDeepConvolutionalLSTM2016}. The DeepConvLSTM still remains one of the most popular and effective architectures in the field and was thus chosen by us to be architecture of choice. Nevertheless, in later steps of the tutorial, we will also demonstrate participants the modularity of the DL-ARC by showing how one would interchange the architecture to be applied by the pipeline.

\vspace{0.1cm}

\begin{figure}[h]
  \centering
  \includegraphics[width = \linewidth]{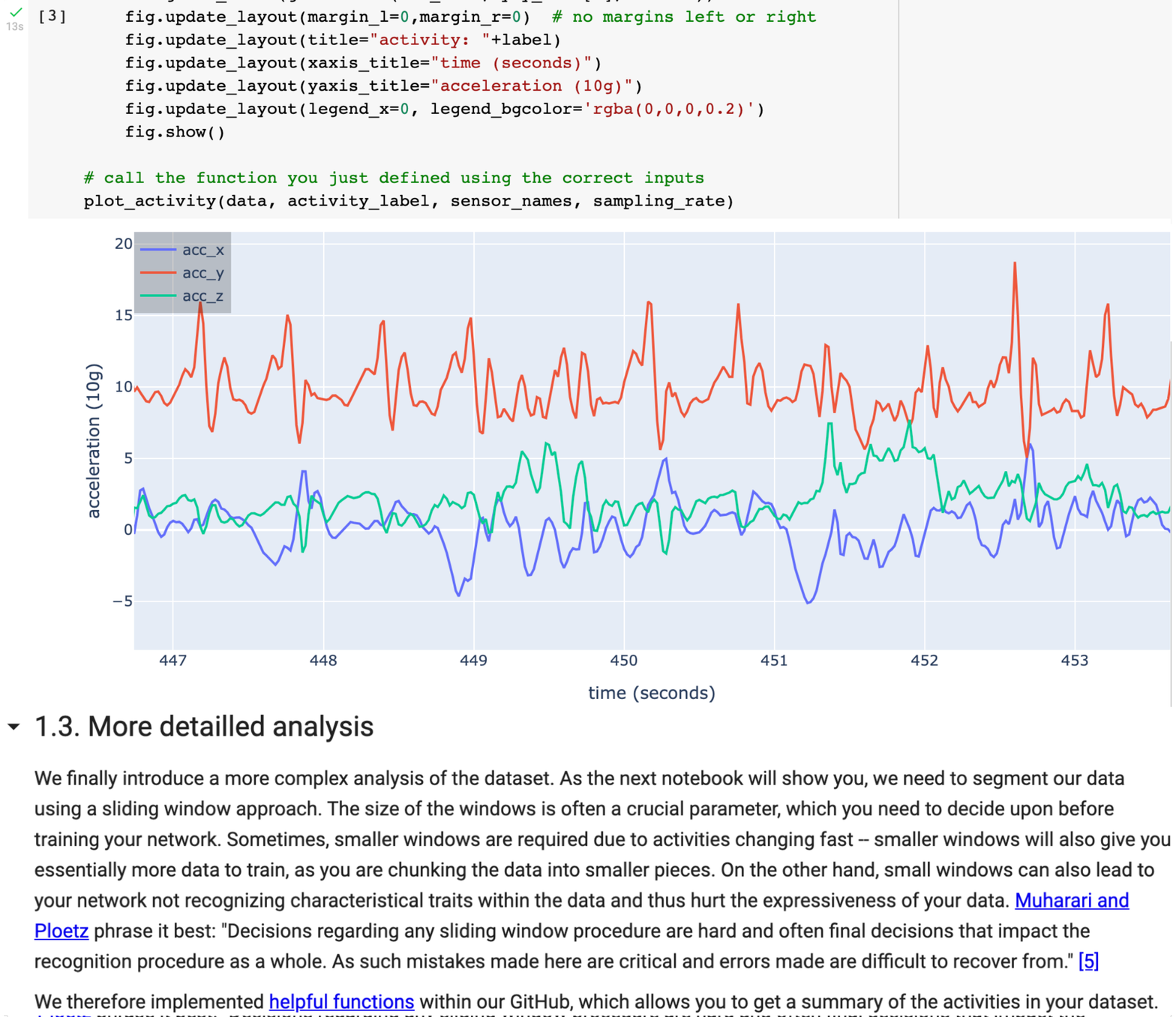}
  \caption{A screenshot of one of the interactive Colab pages, where blocks of python code explain basic steps in Jupyter notebooks, for instance for visualization, accompanied by text blocks providing background explanations.}
  \Description{A screenshot of one of the interactive Colab pages, where blocks of python code explain basic steps in Jupyter notebooks, for instance for visualization, accompanied by text blocks providing background explanations.}
  \label{fig:shot}
\end{figure}

\noindent \textbf{The DL-ARC pipeline \ \ } The second phase introduces participants to the DL-ARC pipeline. Using a mix of theoretical session and hands-on coding exercises, participants are guided step-by-step through each of the four main parts which the pipeline consists of. The hands-on exercises are presented as Jupyter notebooks, which can be accessed by the participants through Google Colab or locally on their workstation (see Figure \ref{fig:shot}). We recommend participants to use former as Google Colab already offers a working Python distribution as well as access to GPU resources, and therefore requires no setup effort in order to run our code.

\paragraph{Data collection and analysis} Within the first step of the DL-ARC pipeline we give a brief overview of challenges faced with when collecting a sensor-based (HAR) dataset. We briefly discuss the different ways of collecting sensor-based data as well as challenges one faces when collecting a dataset. We further explain how accelerometers work as well as explain fundamental terms like sampling rate, interval and range. Afterwards we introduce the dataset which is used throughout the tutorial, namely the RealWorld HAR (RWHAR) dataset \cite{sztylerOnBodyLocalizationWearable2016}. The dataset contains data of 15 participants performing 8 different activities (\textit{walking upstairs, walking downstairs, jumping, lying, standing, sitting, running, walking}). To account for the limited timeframe of the tutorial, we will have participants use a smaller version of the dataset in order to have shorter training times. Within the first coding exercise participants are then asked to familiarise themselves with how one can load the RWHAR dataset as well as apply sample data analysis steps. This includes ways to analyse the activities performed in the dataset (e.g. average length, label distribution) and visualization techniques of the sensor streams.

\paragraph{Preprocessing} Within the second part of the DL-ARC pipeline participants are taught everything related to preprocessing HAR datasets. In order to do so we first name frequent issues that arise when dealing with sensor-based datasets like missing values, different sampling rates and sensitivities and then name sample methods which can be used to solve said problems, namely interpolation, resampling and scaling. Furthermore, we go into detail why it is necessary to segment sensor-data into sliding windows. To conclude the chapter the related coding exercise then asks participants to apply said preprocessing methods as well as also highlights pitfalls that frequently arise when applying them.

\paragraph{Network architecture and training} The third part functions as an introduction to deep learning and teach participants fundamental terms and concepts surrounding it. Before talking about which parts a deep learning training loop consists of we clarify what the term deep learning actually means. Moreover, this part also covers sample evaluation metrics that can be applied to judge the predictive performance of a classifier. Though it was not covered in the initial version of this tutorial, we offer participants to work through a separate, additional Jupyter notebook, which goes into more detail about how these metrics are computed using a toy classification example. We then talk about the types of layers one can find in a sample deep learning as well as their purpose and function, and introduce a the architecture which is used during the tutorial, namely an altered version of the DeepConvLSTM \cite{ordonezDeepConvolutionalLSTM2016}, which was introduced in recent publication of ours \cite{bockImprovingDeepLearning2021}. During the coding exercise related to his chapter participants are asked to implement the architecture introduced to them using PyTorch as well as write a sample training loop. 

\paragraph{Validation and testing} Within the last part of introducing the DL-ARC pipeline, participants are taught about the importance of validation and testing. The fundamental pitfall that is overfitting is explained as well as shown how to be avoided via three sample validation methods, namely train-validation splits, k-fold and cross-participant cross-validation and stressing the necessity of testing. Additionally, we briefly speak about hypertuning and how it should be approached as well as partly "automatized" by using frameworks like the BOHB framework \cite{falknerBOHBRobustEfficient2018}. To conclude the second phase, as part of the coding exercise, participants work on implementing the validation methods themselves and test their trained networks.

\vspace{0.1cm}
\noindent \textbf{Next research steps \ \ } 
After demonstrating the fundamentals of the DL-ARC, the last part of the tutorial presents some opportunities of future research by exploiting some exemplary advances in the field of deep learning. The goal of this part is to give some inspiration for extending the DL-ARC in order to possibly improve results. We discuss two computer vision techniques, which - to the best of our knowledge - have not been exploited for HAR yet, namely improving generalization properties using MaxUp \cite{gongMaxUpLightweightAdversarial2021} and incorporating a level of uncertainty into the training by using label smoothing as proposed in \cite{szegedyRethinkingInceptionArchitecture2016} and, for instance, analyzed in \cite{NeurIPS19Mueller}. While the limited time of the tutorial did not allow to implement advanced techniques live, we hope to inspire participants to conduct research on HAR network architectures and training schemes with the motivation to beat the preliminary results they obtained in our tutorial. To foster such research we provide additional resources, references and a coding exercise surrounding the discussed techniques in the form of a final Jupyter notebook. All in all, we hope that this lays the foundation for participants to utilize, test and extend all the material provided during the tutorial far beyond the day of the tutorial. 




\section{Support materials}
Participants are only required to bring their own laptop to the tutorial. As training a neural networks requires GPU resources we recommend participants to use Google Colab, which is tightly integrated with GitHub. The tutorial's presentation slides, which contain links to all Colab websites with Jupyter Notebooks, can be accessed via:\newline \textbf{\url{https://mariusbock.github.io/dl-for-har}}



\bibliographystyle{ACM-Reference-Format}
\bibliography{main}

\appendix

\end{document}